\documentclass[aps,prb,twocolumn,showpacs,amsmath,amssymb]{revtex4}
\usepackage{graphicx}
\usepackage{epsfig}
\usepackage{dcolumn}
\usepackage{bm}

\begin{document}
\title{Quantum state detection of a superconducting flux qubit using a DC-SQUID in the inductive mode}
\author{A. Lupa\c scu}
\affiliation{Kavli Institute of Nanoscience, Delft University of
Technology, PO Box 5046, 2600 GA Delft, The Netherlands}
\author{C. J. P. M. Harmans}
\affiliation{Kavli Institute of Nanoscience, Delft University of
Technology, PO Box 5046, 2600 GA Delft, The Netherlands}
\author{J. E. Mooij}
\affiliation{Kavli Institute of Nanoscience, Delft University of
Technology, PO Box 5046, 2600 GA Delft, The Netherlands}

\date{\today}

\begin{abstract}
We present a readout method for superconducting flux qubits. The qubit quantum flux state can be measured by determining the Josephson inductance of an inductively coupled DC superconducting quantum interference device (DC-SQUID). We determine the response function of the DC-SQUID and its back-action on the qubit during measurement. Due to driving, the qubit energy relaxation rate depends on the spectral density of the measurement circuit noise at sum and difference frequencies of the qubit Larmor frequency and SQUID driving frequency. The qubit dephasing rate is proportional to the spectral density of circuit noise at the SQUID driving frequency. These features of the backaction are qualitatively different from the case when the SQUID is used in the usual switching mode. For a particular type of readout circuit with feasible parameters we find that single shot readout of a superconducting flux qubit is possible.    
\end{abstract}


\maketitle

\section{Introduction}

An information processor based on a quantum mechanical system can be used to solve certain problems significantly faster than a classical computer~\cite{nielsen_2000_1}. This idea has motivated intense research in recent years on the control and measurement of quantum mechanical systems. The basic units in a quantum computer are two level systems, also called quantum bits or qubits. Many types of qubits based on various physical systems have been proposed and implemented experimentally.

Qubits based on solid state systems have the advantage of flexibility in design parameters and scalability. An important class of solid state qubits are the superconducting qubits. They are mesoscopic systems formed of superconductor structures containing Josephson junctions. The energy level structure in these systems is the result of the interplay between the charging energy, associated with the electrostatic energy due to distribution of the charge of a single Cooper pair, and of the Josephson energy, associated with the tunnelling probability for Cooper pairs across the Josephson junctions. Quantum coherent oscillations have been observed for a few versions of qubits with Josephson junctions~\cite{nakamura_1999_1,vion_2002_1,yu_2002_1,martinis_2002_1,duty_2004_1,chiorescu_2003_1} and coupling of two qubits was demonstrated ~\cite{pashkin_2003_1}.

A suitable qubit state detection apparatus for individual qubits is an essential ingredient for the implementation of algorithms for a quantum computer. Efficient measurement is necessary to extract all the relevant information on single qubit states within a restricted time. Moreover, for correlation type measurements in a multiple qubit system, the unwanted backaction of the first measurement should not disturb the system so strongly, that subsequent measurements will be meaningless. In this paper we discuss a measurement method for superconducting flux qubits. Flux qubits are a qubit variety formed of a superconducting loop interrupted by Josephson junctions. The basis states have oppositely circulating persistent currents in the loop. The control parameter is an external magnetic flux in the qubit loop. The qubit state can be determined by measuring the magnetic flux generated by its persistent current.

A natural candidate for the measurement of the state of a flux qubit is a DC superconducting quantum interference device (DC-SQUID). A DC-SQUID is a loop containing two Josephson junctions. Its critical current, which is the maximum supercurrent that it can sustain, depends on the magnetic flux enclosed in the loop~\cite{barone_1982_1}. The state of a flux qubit~\cite{mooij_1999_1,orlando_1999_1}, was measured using an underdamped DC-SQUID~\cite{vanderwal_2000_1,chiorescu_2003_1}. The critical current of the SQUID, and thus the state of the flux qubit, is determined as the maximum value of the current, where the SQUID switches to a finite voltage state. Due to thermal and quantum fluctuations, switching of the SQUID is a stochastic process~\cite{devoret_1985_1}. The qubit states are distinguishable if the difference between the two average values of the switching current, corresponding to the qubit flux states, is larger than the statistical spread of the measured values of the switching current. The measurement of a flux qubit using a switching DC-SQUID was characterized by an efficiency as large as 60~$\%$~\cite{chiorescu_2003_1}. Further improvement of the measurement efficiency is possible and the backaction on the measured qubit if no measurement is performed can be reduced to acceptable levels~\cite{vanderwal_2003_1}. Nevertheless, switching to the dissipative state has a few drawbacks. The finite voltage across the DC-SQUID determines the generation of quasiparticles which causes decoherence of the qubit~\cite{lang_2003_1}. The long quasiparticle recombination time is a severe limit to the reset times for the qubits. In the finite voltage state the SQUID generates AC signals with frequencies in the microwaves range and broad spectral content, that can induce transitions in a multiple qubit system, constrained to have energy levels spacings in the same region. The mentioned types of backaction will not have an affect on the statistics of the measurements on a \emph{single} qubit; however, in a complex multiple qubit system switching of a DC-SQUID to the finite voltage state is a strong disturbance of the state of the total system which introduces errors in subsequent computations and/or measurements.

A DC-SQUID can be used as a flux detector in an alternative mode of operation, in which switching to the dissipative state is avoided. This is based on the property of a SQUID to behave as an inductor, with a Josephson inductance that depends on the magnetic flux enclosed in the loop~\cite{barone_1982_1}. The value of the Josephson inductance can be determined by measuring the impedance of the SQUID. The flux sensitivity in this operation mode is increased if the junction is shunted by a capacitor and the circuit is excited with an AC signal at a frequency close to the resonance frequency. A SQUID in the inductive mode integrated in a resonant circuit was used for the measurement of spectroscopy of a flux qubit~\cite{lupascu_2004_1}.

The inductive operation mode resembles the RF-SQUID in the dispersive mode~\cite{barone_1982_1}. The RF-SQUID contains a superconducting loop with a single Josephson junction; the impedance of a high quality tank circuit inductively coupled to the loop is measured near resonance, where it is very sensitive to the value of the magnetic flux in the loop. For charge measurement, a similar device is the RF single electron tunnelling transistor (RF-SET), with the difference that a dissipative property of a SET transistor is measured directly. The RF-SET was used as a detector for charge qubits by Duty~\emph{et al.}~\cite{duty_2004_1}. Motivated by research on superconducting qubits a few flux or charge detectors based on the measurement of a reactive circuit element have been recently implemented. A flux qubit was studied by Grajcar~\emph{et al.}~\cite{grajcar_2004_1} by measuring the susceptibility of the qubit loop using a coupled high quality tank circuit. A detector for charge qubits based on the measurement of the inductance of a superconducting SET was proposed by Zorin~\cite{zorin_2001_1} and implemented by Sillanp\"{a}\"{a}~\emph{et al.}~\cite{sillanpaa_2004_1}. A sensitive measurement of the critical current of a Josephson junction which exploits the nonlinearity of the current phase relation was demonstrated by Siddiqi~\emph{et al.}~\cite{siddiqi_2003_2}. The state of a charge qubit was read out by Wallraff~\emph{et al.}~\cite{wallraff_2004_1} by measuring the transmission through a coupled transmission line resonator. 

The paper is organized as follows. In Sec. \ref{sec:considerations} we discuss a few general constraints on the parameter range where the DC-SQUID in the inductive mode can operate. We continue in Sec. \ref{sec:response} with a general analysis of the response function of this device as a flux detector. The response function is derived for a general type of circuit embedding the DC-SQUID. In Sec. \ref{sec:interaction} we discuss the qubit-SQUID interaction and we identify the relevant aspects of the measurement backaction. The energy relaxation rate and the dephasing rate of the qubit during the measurement are derived in Sec.~\ref{sec:decoherence}. Because of AC driving of the SQUID and quadratic coupling of the qubit to the SQUID, the qubit relaxation rate is proportional to the spectral density of circuit noise at frequencies which are the sum and the difference of SQUID AC driving frequency and qubit Larmor frequency. Similarly, the dephasing rate is proportional to the spectral density of circuit noise at the frequency of the SQUID AC driving. In Sec.~\ref{sec:discussion} we discuss the results of the calculations on the measurement backaction. We analyze the measurement efficiency for a specific readout circuit and we find that single shot readout of a flux qubit is possible.

\section{General considerations}
\label{sec:considerations}

\begin{figure}[!]
\includegraphics[width=3.4in]{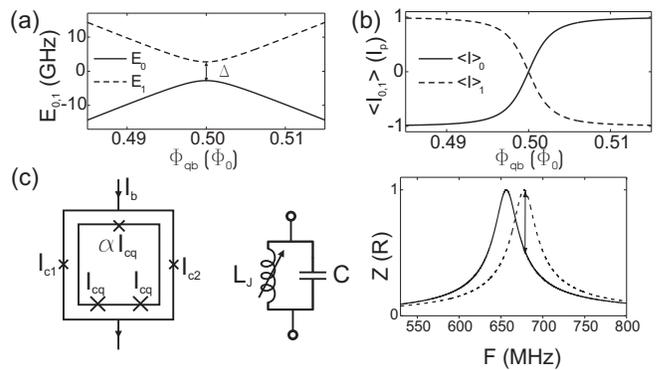}
\caption{\label{fig1} (a) Ground and excited state energy levels for a PCQ, with $E_{J} = 258$ GHz, $E_{C} = 6.9$ GHz and $\alpha = 0.75$, which results in $I_{p} = 300$ nA and $\Delta = 5.5$ GHz. (b) Expectation value of the loop current for the ground and excited energy states with the parameters mentioned in (a). (c) Schematic representation of the PCQ and of the measuring DC-SQUID, with crosses indicating Josephson junctions. The SQUID acts like a variable inductor, with a value dependent on the qubit flux state. The impedance of a resonant circuit formed of the DC-SQUID and a shunt capacitor depends on the qubit state (R represents an equivalent series resistance for the resonance circuit, not shown).}
\end{figure}

In this paper we focus on the readout of a persistent current qubit (PCQ)~\cite{mooij_1999_1,orlando_1999_1}, though the analysis of the measurement is applicable to flux qubits in general. The PCQ is formed of a superconducting loop with three Josephson junctions. Two of the three junctions are of equal size, with Josephson energy $E_{J}$ and charging energy $E_{c}$, while the third junction is smaller by a factor $\alpha$. Figure~\ref{fig1}a shows a representation of the energy levels versus the value of the external magnetic flux in the loop for a set of typical parameters. The qubit quantum state can be represented as a superposition of two basis states that are persistent current states in the loop, with opposite values. Away from the symmetry point $\Phi_{qb}=\Phi_{0}/2$ the energy eigenstates are almost equal to the persistent current states. When $\Phi_{qb}$ approaches $\Phi_{0}/2$, the energy eigenstates are superpositions of the basis states and for $\Phi_{qb}=\Phi_{0}/2$ they become the symmetric and antisymmetric combinations of the basis persistent current states. A representation of the expectation value of the current in each energy eigenstate is given in Fig.~\ref{fig1}b.

The DC-SQUID is characterized by the gauge-invariant phase variables across the two Josephson junctions, denoted by $\gamma_{1}$ and $\gamma_{2}$. The two variables are connected through the fluxoid quantization condition, $\gamma_{1}-\gamma_{2} = -2\pi\Phi_{sq}/\Phi_{0}$, where $\Phi_{0}$ is the flux quantum and $\Phi_{sq}$ is the total flux in the SQUID loop. The flux in the SQUID loop contains an external component $\Phi_{x}$ and a self generated component, which can be neglected for the typical parameters we will discuss. With this assumption, $\gamma_{1}-\gamma_{2} = -2\pi f_{x}$, where $f_{x} = \Phi_{x}/\Phi_{0}$ and the SQUID can be described as a single Josephson junction with a critical current given by $I_{c}\left(f_{x}\right) = 2I_{c0}|\cos\left(\pi f_{x}\right)|$ for a symmetric SQUID ($I_{c1} = I_{c2} = I_{c0}$). For symmetric qubit-SQUID coupling (when the two SQUID branches have mutual inductances to the qubit loop with opposite value) the flux generated by the SQUID in the qubit loop is given by $M_{s}I_{circ}$, where $M_{s}$ is the inductance between the qubit and the SQUID loops and $I_{circ} = \left(I_{1}-I_{2}\right)/2$ is the circulating current in the SQUID loop. The latter is given by 
\begin{equation}\label{eq:Icirc}
I_{circ} = I_{c0}\sin\left(\pi f_{x}\right)\cos{\gamma_{e}},
\end{equation}  
where $\gamma_{e} = \left(\gamma_{1}+\gamma_{2}\right)/2$.
The current and the voltage of the SQUID are related to the variable $\gamma_{e}$ through the two Josephson relations:
\begin{equation}\label{eq:Josephson1}
I = I_{c}\left(f_{x}\right)\sin{\gamma_{e}},
\end{equation} 
and
\begin{equation}\label{eq:Josephson2}
V = \frac{\Phi_{0}}{2\pi}\frac{d\gamma_{e}}{dt}.
\end{equation}
From Eq.~\ref{eq:Josephson1} and \ref{eq:Josephson2} it follows that in the linear approximation ($\sin\left(\gamma_{e}\right) \simeq \gamma_{e}$) the SQUID behaves as a linear inductor with the Josephson inductance
\begin{equation}\label{eq:Josephson_inductance}
L_{J} = \frac{\Phi_{0}}{2\pi I_{c}}.
\end{equation}

If an AC current is injected in the SQUID at frequency $\nu_{0}$ with a small amplitude $I_{ac}$, the voltage across the SQUID has the amplitude $V = 2\pi\nu_{0}L_{J}I_{ac}=\Phi_{0}\nu_{0} I_{ac}/I_{c}$. The maximum voltage across the SQUID is very small ($\sim\: 2 \mu$V for $\nu_{0}=1$ GHz); very low noise amplification is necessary to detect such a voltage in a short time. Increasing the value of $\nu_{0}$ will result in a proportional increase in the value of the maximum AC voltage. However, from Eq.~\ref{eq:Icirc} and \ref{eq:Josephson1} it follows that when the SQUID current varies at frequency $\nu_{0}$, the circulating current contains a significant frequency component at $2\nu_{0}$ and additional higher harmonics for strong driving in the nonlinear regime. The flux generated by this circulating current in the qubit loop can cause transitions between the qubit energy levels if the harmonics of the driving current are close to the qubit energy levels splitting. With typical level splitting of $1-20$ GHz, the value of $\nu_{0}$ is limited to $\lesssim 1$ GHz.

The relative change in Josephson inductance when the qubit evolves from the ground to the excited state is given by
\begin{equation}\label{eq:relative_change_Josephson}
\left|\frac{\delta L_{J}}{L_{J}}\right|\simeq \left|\frac{\delta I_{c}}{I_{c}}\right|\simeq 2 \pi |\tan (\pi f_{x})|\frac{M_{s}I_{p}}{\Phi_{0}},
\end{equation}
where it was assumed that the measurement is performed at a bias flux in the qubit away from $\Phi_{0}/2$, so that the expectation value of the qubit current in each energy eigenstate approaches in absolute value $I_{p}$ - the maximum value of the persistent current (see Fig.~\ref{fig1}b). The typical values for $M_{s}$ and $I_{p}$ limit the value of $\delta L_{J}/L_{J}$ to a few percent. If the SQUID is driven with a constant AC current, the maximum difference in AC voltage corresponding to a qubit state change, from the ground to the excited state, is $\sim \Phi_{0} \nu_{0} \delta L_{J}/L_{J}$. This can be increased if the SQUID is placed in a resonant circuit and the frequency $\nu_{0}$ is taken close to the circuit resonance frequency (see Fig.~\ref{fig1}c). A limit on the quality factor $Q$ will be set by the fact that the response time of the resonator, $Q/\omega_{0}$, has to be smaller than the intrinsic qubit relaxation time, which is in the microseconds range~\cite{chiorescu_2003_1,lupascu_2004_1}. When $Q > 2 L_{J}/\delta L_{J}$ the two circuit resonance peaks, corresponding to the different qubit states, are separated and a further increase of $Q$ will not contribute to an increase in the AC voltage difference.

The above considerations show that, given the typical qubit energy level splitting and relaxation time, the constraints on the circuit parameters are $\nu_{0} \lesssim 1 $GHz and $Q < 100$.

\section{Detector response function}
\label{sec:response}
In this section we analyze the DC-SQUID in the inductive mode as a flux detector. We consider the case of moderate AC driving, when the SQUID behaves as a linear inductor. The function describing the conversion of flux in the SQUID loop to AC voltage is determined for a general type of circuit in which the SQUID is embedded.

\begin{figure*}[!]
\includegraphics[width=6.0in]{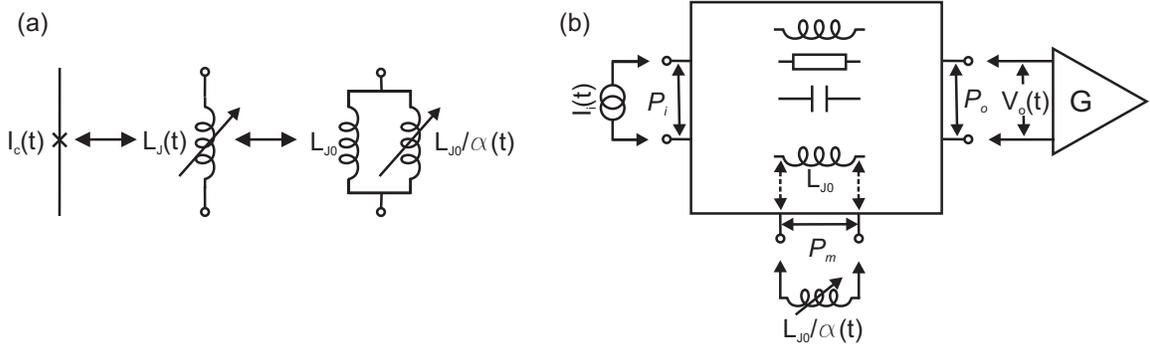}
\caption{\label{fig2} (a) The time dependent Josephson inductance can be represented as a series combination of the inductors $L_{J0}$ and $L_{J0}/\alpha(t)$. (b) Schematic representation of the circuit in which the DC-SQUID is inserted (see the text for explanations). }
\end{figure*}

If the magnetic flux in the SQUID loop varies in time, the relation between the transport current and the voltage across the terminals of the SQUID is given by
\begin{equation}\label{eq:IV_inductor}
I(t)=\frac{1}{L_{J}(t)}\int^{t} V(t')dt',
\end{equation}
where $L_{J}\left(t\right)$ is the time dependent Josephson inductance. Let us consider 
\begin{equation}\label{eq:Jos_inductance_parametrize}
\frac{1}{L_{J}(t)}=\frac{1}{L_{J0}}(1+\alpha (t)),
\end{equation}
where $\alpha\left(t\right)$ parameterizes the variations of the magnetic flux in the SQUID. The time dependent Josephson inductance $L_{J}(t)$ can be represented as the parallel combination of the inductances $L_{J0}$ and $L_{J0}/\alpha(t)$ (see Fig.~\ref{fig2}a). In the case of the qubit measurement, $\alpha (t)$ describes the dynamics of the qubit generated flux.  The extreme values of $\alpha (t)$ in this case correspond to the qubit in a clockwise or anticlockwise persistent current state and are given approximately by $\pm \pi \tan (\pi f_{x})MI_{p}/\Phi_{0}$ (see Eq.~\ref{eq:relative_change_Josephson}). We assume $|\alpha\left(t\right)| << 1$, consistent with the usual value of the flux generated by the coupled qubit in the SQUID loop which is of the order of $1\%$ of $\Phi_{0}$ ~\cite{vanderwal_2000_1,chiorescu_2003_1,lupascu_2004_1}. From Eq.~\ref{eq:IV_inductor} and~\ref{eq:Jos_inductance_parametrize}, it follows that the current in the SQUID can be written as 
\begin{equation}\label{eq:current_separation}
I=I_{0}+I_{1},
\end{equation}
with 
\begin{equation}\label{eq:big_current}
I_{0}(t)=\frac{1}{L_{J0}}\int^{t} V(t')dt'
\end{equation}
and
\begin{equation}\label{eq:small_current}
I_{1}(t)=\frac{\alpha(t)}{L_{J0}}\int^{t} V(t')dt'.
\end{equation}

As discussed in section~\ref{sec:considerations}, the measurement of the Josephson inductance is more efficient if the DC-SQUID is integrated in a resonant circuit. In this case the circuit is driven with an AC current source at a frequency close to the resonance frequency and the AC voltage is suitably amplified. The output AC voltage depends on $\alpha (t)$. Since there is a certain freedom in the design of the resonant circuit, we calculate here the dependence of the output voltage on $\alpha(t)$ for a general type of circuit in which the DC-SQUID is embedded. We consider a linear network with three ports that contains $L_{J0}$ (the constant component of the SQUID inductance), the impedance of the driving current source, the amplifier input impedance, complemented by other linear circuit elements. The port $P_{i}$ has its terminals across the current source. The port $P_{m}$ is connected across the Josephson inductance $L_{J0}$. Finally, the port $P_{o}$ has its connections at the input of the amplifier. Three elements are connected to the ports $P_{i}$, $P_{o}$ and $P_{m}$, respectively: the ideal current source $I_{i}(t)$, an ideal voltage amplifier and the inductance $L_{J0}/\alpha (t)$. A specific electrical circuit described in the way indicated here is shown in Fig.~\ref{fig3}a. 

The relation between the current and the voltage at port $P_{m}$ is determined by the inductance $L_{J0}/\alpha (t)$:
\begin{equation}\label{eq:branch_m}
-I_{m}(t)=\frac{\alpha(t)}{L_{J0}}\int^{t} V_{m}(t')dt'
\end{equation}
which is Eq.~\ref{eq:small_current} with changed sign in order to preserve the sign convention for the 3-ports network.
The voltage at the port $P_{\alpha}$, $\alpha=o,m$ can be written as 
\begin{equation}\label{eq:port_o_and_m}
V_{\alpha}(t)=\int_{0}^{\infty}Z_{\alpha i}(t')I_{i}(t-t')dt'+\int_{0}^{\infty}Z_{\alpha m}(t')I_{m}(t-t')dt'.
\end{equation}
Here $Z_{\alpha\beta}(t)$ is the impedance matrix for the 3 port network, with $\alpha,\beta=i,m,o$. We assume that the AC driving current is $I_{i}(t)=I_{e}\cos(\omega_{0}t)$ and the flux variations in the DC-SQUID loop (see Eq.~\ref{eq:Jos_inductance_parametrize}) are described by $\alpha(t) = Re (\alpha_{0}\exp(-i\omega t))$. Equation~\ref{eq:branch_m} implies that $V_{m}$ and $I_{m}$ have components at frequencies $\omega_{0}+n\omega$, with $n$ being an integer. 
The voltage $V_m$ can be written as:
\begin{equation}\label{eq:sum_voltage_port_m}
V_{m}(t)=\sum_{n}(\frac{V_{m,n}}{2}e^{-i(\omega_{0}+n\omega) t}+\frac{V_{m,n}^{*}}{2}e^{i(\omega_{0}+n\omega) t}).
\end{equation}
A similar expression for $I_{m}$ can be written if $V$ is replaced by $I$ in expression~\ref{eq:sum_voltage_port_m}.

From Eq.~\ref{eq:port_o_and_m} written for $\alpha=m$, one obtains the Fourier components of $V_{m}(t)$ as a function of the Fourier coefficients of $I_{m}(t)$ (note that $I_{i}(t)$ is imposed and has frequency components at $\pm \omega_{0}$). These values can be replaced in \ref{eq:branch_m}, and the terms corresponding to the frequencies $\omega_{0}+n\omega$ are separated. Using the equations corresponding to $n=0,1,-1$ in expansions of the form \ref{eq:sum_voltage_port_m} and neglecting the terms $I_{m,2}$ and $I_{m,-2}$ , the values for $I_{m,0}$, $I_{m,1}$ and $I_{m,-1}$ can be obtained in lowest order in $|\alpha_{0}|$. Using these values in Eq. \ref{eq:port_o_and_m} for $\beta=o$ leads to the following expression for the components of the output voltage at frequencies $\omega_{0}+\omega$ and $\omega_{0}-\omega$:
\begin{equation}\label{eq:voltage_up_converted}
V_{o,1}=\frac{\alpha_{0}}{i\omega_{0}L_{J0}}Z_{mi}(\omega_{0})\frac{I_{e}}{2}Z_{om}(\omega_{0}+\omega)
\end{equation}
and
\begin{equation}\label{eq:voltage_down_converted}
V_{o,-1}=\frac{\alpha_{0}^{*}}{i\omega_{0}L_{J0}}Z_{mi}(\omega_{0})\frac{I_{e}}{2}Z_{om}(\omega_{0}-\omega)
\end{equation}

The expressions \ref{eq:voltage_up_converted} and \ref{eq:voltage_down_converted} for the up- and down-converted voltage at the output of the circuit are proportional to the amplitude of the driving current $I_{e}$ and to the amplitude of the flux modulation $|\alpha_{0}|$. These expressions are usable only when the driving conditions ($I_{e}$ and $\omega_{0}$) are such that the maximum amplitude of the current in the SQUID is not close to the SQUID critical current. Besides the up- and down- converted components $V_{o,1}$ and $V_{o,-1}$, the output voltage contains a strong component $V_{o,0}$ at frequency $\omega_{o}$. $V_{o}$ depends only quadratically on $|\alpha_{0}|$ and thus it cannot be used for an efficient detection of the flux.

\section{Qubit-SQUID interaction}
\label{sec:interaction}
In a basis formed of two persistent current states, the Hamiltonian of the flux qubit can be written as~\cite{orlando_1999_1}
\begin{equation}\label{eq:qubit_Hamiltonian}
\hat{H}_{qb}=\frac{\epsilon}{2}\hat{\sigma}_{z}+\frac{\Delta}{2}\hat{\sigma}_{x},
\end{equation}
where $\hat{\sigma}_{i},\ i=x,y,z$, have the Pauli matrices representation. The coefficient of the first term in~\ref{eq:qubit_Hamiltonian} is $\epsilon=2I_{p}\left(\Phi_{qb}-\Phi_{0}/2\right)$, where $\Phi_{qb}$ is the flux in the qubit loop. The maximum persistent current $I_{p}$ and the minimum energy level splitting $\Delta$ (see Fig.~\ref{fig1}a and b) are parameters fixed by the qubit junctions design. The average flux induced in the SQUID loop by the qubit is $MI_{p}\langle\hat{\sigma}_{z}\rangle$. The flux dependent term in the energy of a SQUID is the Josephson energy given by $-2\Phi_{0}/(2\pi)I_{c0}\cos(\gamma_{e})\cos(\pi\Phi_{sq}/\Phi_{0})$. The total flux $\Phi_{sq}$ in the SQUID loop contains the external flux $\Phi_{x}$ and qubit induced flux $MI_{p}\langle\hat{\sigma}_{z}\rangle$. It follows that the interaction Hamiltonian can be written as
\begin{equation}\label{eq:coupling_Hamiltonian}
\hat{H}_{c}=MI_{p}I_{c0}\cos(\hat{\gamma}_{e})\sin(\pi f_{x})\hat{\sigma}_{z},
\end{equation}
where we assumed that the flux generated by the qubit is small and thus a linear approximation could be used. A rigorous derivation of the interaction term in the Hamiltonian for a coupled DC-SQUID and a three Josephson junctions qubit, assuming a SQUID with a small self inductance and using the two level approximation for the 3 Josephson junctions loop leads to the same result as~\ref{eq:coupling_Hamiltonian}. 

If ~\ref{eq:coupling_Hamiltonian} is compared to \ref{eq:qubit_Hamiltonian} with $\epsilon/2 = I_{p}\left(\Phi_{qb}-\Phi_{0}/2\right)$, it becomes clear that the back-action due to the measurement is described by an equivalent flux operator, expressed as $\hat{\Phi}_{b} = M\hat{I}_{circ}$. $\hat{I}_{circ}$ is the operator corresponding to the circulating current in the DC-SQUID and is given by $\hat{I}_{circ}=I_{f}\cos(\hat{\gamma_{e}})$, where $I_{f}=I_{c0}\sin(\pi f_{x})$ (see Eq.~\ref{eq:Icirc}). Classically, qubit decoherence can be understood as a result of the fluctuations in the flux bias, due to the SQUID. If $\gamma_{e}$ is treated as a classical variable, its time evolution is given by 
\begin{equation}\label{eq:ext_phase_classical_picture}
\gamma_{e}(t)=\gamma_{e,coh}(t)+\gamma_{e,n}(t)
\end{equation}
where $\gamma_{e,coh}(t)=Re\left(\gamma_{e0}\exp\left(-i\omega_{0}t\right)\right)$ with $\gamma_{e0}=I_{e}Z_{im}\left(\omega_{0}\right)/\left(i\omega_{0}\right)$ the response to circuit driving and $\gamma_{e,n}(t)$ is a random term, corresponding to e.g. thermal fluctuations. The "classical" flux is given by $\Phi_{b}(t) = MI_{f}\cos(\gamma_{e}(t))$, which can be approximated by 
\begin{equation}\label{eq:backaction_flux_classical}
\Phi_{b}(t) = MI_{f}\left(1-\frac{\gamma_{e}^{2}\left( t\right)}{2}\right).
\end{equation}
The statistical properties of $\Phi_{b}\left( t\right)$ are thus determined by $\gamma_{e}^2\left( t\right)$. From~\ref{eq:ext_phase_classical_picture} it follows that for the case when the phase oscillations amplitude is large compared to the typical values of $\gamma_{en}\left( t\right)$, the most important contribution will be the mixing term $\gamma_{e,coh}(t)\gamma_{e,n}(t)$. This results in frequency conversion of the circuit noise. 

The analog of equation \ref{eq:ext_phase_classical_picture} for the quantized system is
\begin{equation}\label{eq:ext_phase_interaction_picture}
\hat{\gamma}_{e}^{I}(t)=\gamma_{e,coh}(t)+\hat{\gamma}_{en}(t)
\end{equation}
in which $\hat{\gamma}_{e}^{I}(t)$ is the phase operator in the interaction representation \emph{with respect to the qubit-SQUID interaction}, which is thus equivalent to the Heisenberg representation for the SQUID system. $\hat{\gamma}_{e,n}(t)$ is the phase operator representing the intrinsic evolution, in the absence of circuit driving. The first term on the right hand side of Eq.~\ref{eq:ext_phase_interaction_picture} corresponds to the evolution due to driving, and it is the same as the first term on the right hand side of Eq.~\ref{eq:ext_phase_classical_picture}. 

The essential feature of the interaction Hamiltonian given by \ref{eq:coupling_Hamiltonian} is that the coupling to the external phase operator does not have a linear part. Recent work on the influence of nonlinear coupling of the noise on the evolution of a two level system has been done~\cite{makhlin_2004_1} motivated by results reported by Vion~\emph{et al.}~\cite{vion_2002_1}, where long coherence times were obtained for the operation of a qubit at settings where the energy level separation was insensitive in the first order to external noise. In the second order approximation, the back-action flux noise is described by
\begin{equation}\label{eq:flux_backaction}
\hat{\Phi}_{b} =  MI_{f}\left(1-\frac{\hat{\gamma}_{e}^{2}}{2}\right),
\end{equation}
which can be separated in three parts as:
\begin{equation}\label{eq:flux_backaction_separated}
\hat{\Phi}_{b1} = MI_{f}\left[\left(1-\frac{\gamma_{e,coh}^{2}(t)}{2}\right)-\frac{\hat{\gamma}_{en}^{2}(t)}{2}-\gamma_{e,coh}(t)\hat{\gamma}_{en}(t)\right],
\end{equation}
The first term on the right hand side of \ref{eq:flux_backaction_separated} can cause transitions between the qubit energy levels. As we discussed in Sec~\ref{sec:considerations}, resonant transitions occur when the qubit energy levels splitting is close to the harmonics of the driving AC frequency, and in particular to the second harmonic. The time average of the first term in~\ref{eq:flux_backaction_separated}, dependent on the amplitude of the AC driving current, will be considered a part of the qubit flux bias $\Phi_{qb}$. The effects of the second term in \ref{eq:flux_backaction_separated} were analyzed by Makhlin \emph{et al.}~\cite{makhlin_2004_1} for ohmic and $1/\omega$ type spectral densities. In this paper we focus on the calculation of decoherence determined by the third term in~\ref{eq:flux_backaction_separated}. In the second order perturbation theory, used for the calculation of the decoherence rates in the next section, the contributions from the different terms in ~\ref{eq:flux_backaction_separated} can be treated independently.

\section{Calculation of the decoherence rates}
\label{sec:decoherence}
In this section we calculate the relaxation and dephasing rates of a flux qubit during the measurement by a DC-SQUID in the inductive mode. It is assumed that the external qubit flux $\Phi_{qb}$ is fixed. However, the calculations can be extended to include the case of the measurement performed during induced Rabi oscillations or other control sequences~\cite{martinis_2003_1,smirnov_2003_1}. The model Hamiltonian used for the combined system qubit-SQUID is:
\begin{equation}\label{eq:Hamiltonian}
\hat{H}(t)=\hat{H}_{qb}+\hat{H}_{c}+\hat{H}_{SQUID}(t),
\end{equation}
where $\hat{H}_{qb}$ is the qubit Hamiltonian given by \ref{eq:qubit_Hamiltonian}, $\hat{H}_{c}$ is the interaction term given by \ref{eq:coupling_Hamiltonian} and $\hat{H}_{SQUID}(t)$ is the SQUID Hamiltonian, which is time dependent due to driving.
If a transformation is made to the qubit energy eigenstates, the first two terms in \ref{eq:Hamiltonian} become:
\begin{equation}\label{eq:qubit_Hamiltonian_transformed}
\hat{H}_{qb}=\frac{\sqrt{\epsilon^{2}+\Delta^{2}}}{2}\hat{\tau}_{z}
\end{equation}
and
\begin{equation}\label{eq:interaction_Hamiltonian_transformed}
\hat{H}_{c}=MI_{p}I_{f}\cos(\hat{\gamma}_{e})\left(\cos(\theta)\hat{\tau}_{z}-\sin(\theta)\hat{\tau}_{x}\right),
\end{equation}
where the $\hat{\tau}_{i}$, $i=x,y,z$ are Pauli matrices in the energy eigenstate basis and $\tan(\theta)=\Delta/\epsilon$. In the interaction representation with respect to the qubit-SQUID interaction, the operators $\tau_{i}^{I}$ evolve in time according to 
\begin{equation}\label{eq:coherent_evolution}
\hat{\tau}^{I}_{i}(t)=\mathbf{A}_{ij}(t)\hat{\tau}_{j}
\end{equation}
with the matrix $\mathbf{A}$ given by:
\begin{equation}\label{eq:coherent_evolution_matrix}
\mathbf{A}(t)=\left(%
\begin{array}{ccc}
  \cos(\omega_{01}t) & -\sin(\omega_{01}t) & 0 \\
  \sin(\omega_{01}t) & \cos(\omega_{01}t) & 0 \\
  0 & 0 & 1 \\
\end{array}%
\right)
\end{equation}
in which $\omega_{01}=\sqrt{\epsilon^2+\Delta^2}/\hbar$ is the frequency corresponding to the qubit energy level separation $\sqrt{\epsilon^2+\Delta^2}$.
The evolution of the operators $\hat{\tau}_{i}^{H}$  in the Heisenberg picture is obtained using time-dependent second order perturbation theory. As explained in Section \ref{sec:interaction}, the relevant part of the interaction Hamiltonian (Eq. \ref{eq:interaction_Hamiltonian_transformed}) for the calculation of decoherence is given in the interaction picture by:
\begin{equation}\label{eq:Hint_interaction_picture_transformed}
\hat{H}_{c}^{I}(t)=-MI_{p}I_{f}\gamma_{e,coh}(t)\hat{\gamma}_{e,n}(t)(\cos(\theta)\hat{\tau}_{z}^{I}-\sin(\theta)\hat{\tau}_{x}^{I}).
\end{equation}
The evolution of the operators $\hat{\tau}_{i}^{H}$, which allows to describe the qubit operators expectation values if the initial state is known, is given by:
\begin{equation}\label{eq:perturbation_second_order}
\hat{\tau}_{i}^{H}(t)=\hat{\tau}_{i}^{I}(t)-\frac{1}{\hbar^{2}}\underset{t>t_{1}>t_{2}>0}\iint\left[\left[\hat{\tau}_{i}^{I}(t),\hat{H}_{c}^{I}(t_{1})\right],\hat{H}_{c}^{I}(t_{2})\right].
\end{equation}
In the following, the second term on the right hand side of Eq. \ref{eq:perturbation_second_order} is calculated. For the initial calculation we assume the most general interaction Hamiltonian with linear coupling to the bath, which can be written as 
\begin{equation}\label{eq:general_coupling_Hamiltonian}
\hat{H}_{c}^{I}(t)=\sum_{i=x,y,z}\hat{f}_{i}^{I}(t)\hat{\tau}_{i}^{I}(t),
\end{equation}
where $\hat{f}_{i}$ are bath operators (note that the interaction representation is used in \ref{eq:general_coupling_Hamiltonian}). In the end the form of the operators $\hat{f}_{i}^{I}(t)$ corresponding to our case, as given by Eq.~\ref{eq:Hint_interaction_picture_transformed}, will be considered:
\begin{eqnarray}\label{eq:particular_f_operators}
 \hat{f}_{x}^{I}(t) &=& MI_{p}I_{f}\sin(\theta)\gamma_{e,coh}(t)\hat{\gamma}_{en}(t) \\\nonumber
 \hat{f}_{y}^{I}(t) &=& 0 \\\nonumber
 \hat{f}_{z}^{I}(t) &=& -MI_{p}I_{f}\cos(\theta)\gamma_{e,coh}(t)\hat{\gamma}_{en}(t).
\end{eqnarray}
If the commutation and anti-commutation relations for the $\hat{\tau}_{i}$ operators are used, Eq.~\ref{eq:perturbation_second_order} results in
\begin{widetext}
\begin{equation}\label{eq:perturbation_expression}
\hat{\tau}_{i}^{H}(t)=\underset{j,l,m=x,y,z}\sum \mathbf{A}_{ij}(t)\left(\hat{\tau}_{j}+\frac{2}{\hbar^2}\underset{t>t_{1},t_{2}>0}\iint dt_{1}dt_{2}\left(\hat{\mathbf{O}}_{lj}^{+}(t_{1},t_{2})\hat{\tau}_{l}-\hat{\mathbf{O}}_{ll}^{+}(t_{1},t_{2})\hat{\tau}_{j}-\epsilon_{jlm}\hat{\mathbf{O}}_{lm}^{-}(t_{1},t_{2})\right)\right).
\end{equation}
\end{widetext}
In the last expression: 
\begin{equation}\label{eq:O_operators}
\hat{\mathbf{O}}_{ij}^{\pm}(t_{1},t_{2})=\underset{k,l=x,y,z}\sum\mathbf{A}_{ik}(-t_{1})\hat{\mathbf{C}}_{kl}^{\pm}(t_{1},t_{2})\mathbf{A}_{lj}(t_{2})
\end{equation}
with
\begin{equation}\label{eq:Cplus_operators}
\hat{\mathbf{C}}_{kl}^{+}(t_{1},t_{2})=\frac{1}{2}(\hat{f}_{k}^{I}(t_{1})\hat{f}_{l}^{I}(t_{2})+\hat{f}_{l}^{I}(t_{2})\hat{f}_{k}^{I}(t_{1}))
\end{equation}
and
\begin{equation}\label{eq:Cminus_operators}
\hat{\mathbf{C}}_{kl}^{-}(t_{1},t_{2})=\frac{i}{2}(\hat{f}_{k}^{I}(t_{1})\hat{f}_{l}^{I}(t_{2})-\hat{f}_{l}^{I}(t_{2})\hat{f}_{k}^{I}(t_{1})).
\end{equation}
The last two expressions are symmetrized and anti-symmetrized products of operators at different times. Their expectation values calculated for a thermal equilibrium state are connected with the linear response functions by the fluctuation-dissipation theorem \cite{kubo_1991_1}. Note that, to obtain \ref{eq:perturbation_expression}, the integral in \ref{eq:perturbation_second_order} was extended to the region $t_{1}<t_{2}$  because the integrand is symmetric under the interchange of $t_{1}$ and $t_{2}$. 

We assume that initially the qubit and SQUID states were separable and the SQUID is described by the thermal equilibrium density matrix. For the case of the coupling Hamiltonian given in Eq. \ref{eq:Hint_interaction_picture_transformed}, the relevant correlation functions are:
\begin{equation}\label{eq:sym_corr_function}
C_{\gamma_{e}}^{+}(t_{1},t_{2})=
\langle
\frac{1}{2}
\{
\hat{\gamma}_{e}^{I}(t_{1}),\hat{\gamma}_{e}^{I}(t_{2})
\}_{+}
\rangle_{0}
\end{equation}
and
\begin{equation}\label{eq:asym_corr_function}
C_{\gamma_{e}}^{-}(t_{1},t_{2})=
\langle
\frac{i}{2}
\{
\hat{\gamma}_{e}^{I}(t_{1}),\hat{\gamma}_{e}^{I}(t_{2})
\}_{-}
\rangle_{0},
\end{equation}
where  +/- denote the anticommutator/commutator, and the expectation value is taken for the SQUID thermal equilibrium density matrix. From \ref{eq:perturbation_expression} - \ref{eq:asym_corr_function} we see that the time evolution of the operators $\hat{\tau}_{i}$, $i=x,y,z$, depends on their expectation values for the initial qubit state and on a 2 dimensional integral involving the expectation values of operators of type~\ref{eq:sym_corr_function} and \ref{eq:asym_corr_function}. 

The interaction between the qubit and the measurement DC-SQUID has the consequence that the qubit quantum state becomes a mixed state. In general one distinguishes between \emph{energy relaxation}, corresponding to a change in the qubit energy expectation value, and \emph{dephasing}, corresponding to randomization of the phase of a coherent superposition of energy eigenstates~\cite{martinis_2003_1}. To calculate the energy relaxation, we determine the transition rates between energy eigenstates by determining the evolution of $\langle\hat{\tau}_{z}^{H}(t)\rangle$. If in this calculation the initial qubit state is chosen to be the ground or the excited state, these rates will represent the \emph{absorption} and \emph{emission} rates respectively. To calculate the dephasing rate, we determine the decay of the expectation values $\langle 1/2(\hat{\tau}_{x}^{H}(t)\pm i\hat{\tau}_{y}^{H}(t))\rangle$, with the qubit initial state being a $\hat{\tau}_{x}$ eigenstate.

The calculation of the integral on the right hand side of~\ref{eq:perturbation_expression} involves a product of the functions $\cos(\omega_{01}t_{1,2})$ and $\sin(\omega_{01}t_{1,2})$, resulting from the expression for the free evolution matrix $\mathbf{A}$ (see Eq.~\ref{eq:coherent_evolution_matrix}) and of the functions $\cos(\omega_{0}t_{1,2})$ and $\sin(\omega_{0}t_{1,2})$, resulting from the time dependence of the coupling operators $\hat{f}_{i}^{I}(t)$ (see~\ref{eq:particular_f_operators}). The correlation functions appearing in~\ref{eq:sym_corr_function} and~\ref{eq:asym_corr_function} only depend on the time difference $t_{1}-t_{2}$. For times that are long compared to $2\pi/\omega_{0}$ and $2\pi/\omega_{01}$ the relaxation (decay of $\langle\hat{\tau}_{z}^{H}(t)\rangle$) and the dephasing (decay of $\langle 1/2(\hat{\tau}_{x}^{H}(t)\pm i\hat{\tau}_{y}^{H}(t))\rangle$ ) can be described as an integral of the product of the Fourier transform of one of the spectral functions $C_{\gamma_{e}}^{\pm}(t_{1},t_{2})$ and a weight function that has a width depending on the integration time t. This weight function is given by:
\begin{equation}\label{eq:weight_function}
W(\omega,t)=\frac{1}{2\pi t}\underset{0<t_{1},t_{2}<t}\iint dt_{1}dt_{2}e^{-i\omega(t_{1}-t{2})}
\end{equation}
and has the property $\underset{t\rightarrow\infty}\lim W(\omega,t)=\delta(\omega)$ . The expressions below for the relaxation and dephasing rates are given assuming that the spectral density of the circuit noise, given by the Fourier transform of the correlation functions~\ref{eq:sym_corr_function} and \ref{eq:asym_corr_function}, does not have significant variations over the frequency range where the weight function has substantial values. In this case, the relaxation and the dephasing of the qubit state are proportional to the time from the beginning of the measurement.

The transitions rates between the two energy eigenstates depend on the initial state. The transition rate from the excited state to the ground state $\Gamma_{\downarrow}$ (emission) and the transition rate form the ground state to the excited state $\Gamma_{\uparrow}$ (absorption) are given by:
\begin{equation}\label{eq:emission_rate}
\Gamma_{\downarrow}=\frac{1}{2\hbar^{2}}\sin^{2}(\theta)k^{2}(\gamma_{e0})[S_{\gamma_{e}}(\omega_{01}+\omega_{0})+S_{\gamma_{e}}(\omega_{01}-\omega_{0})]
\end{equation}
and
\begin{equation}\label{eq:absorption_rate}
\Gamma_{\uparrow}=\frac{1}{2\hbar^{2}}\sin^{2}(\theta)k^{2}(\gamma_{e0})[S_{\gamma_{e}}(-\omega_{01}+\omega_{0})+S_{\gamma_{e}}(-\omega_{01}-\omega_{0})]
\end{equation}
in which relations
\begin{equation}\label{eq:coupling_factor}
k\left(\gamma_{e0}\right)=MI_{p}I_{f}|\gamma_{e0}|
\end{equation}
is a measurement coupling factor and 
\begin{equation}\label{eq:spectral_density}
S_{\gamma_{e}}(\omega)=S^{+}_{\gamma_{e}}(\omega)-iS^{-}_{\gamma_{e}}(\omega).
\end{equation}
$S^{\pm}_{\gamma_{e}}$ are the Fourier transforms of the correlation functions $C_{\gamma_{e}}^{\pm}(t,0)$ given by \ref{eq:sym_corr_function} and \ref{eq:asym_corr_function}.
The corresponding expression for the dephasing rate is 
\begin{widetext}
\begin{equation}\label{eq:dephasing_rate}
\Gamma_{\phi}=\frac{1}{4\hbar^{2}}\sin^{2}(\theta)k^{2}(\gamma_{e0})[S^{+}_{\gamma_{e}}(\omega_{01}+\omega_{0})+S^{-}_{\gamma_{e}}(\omega_{01}-\omega_{0})]+\frac{1}{\hbar^{2}}\cos^{2}(\theta)k^{2}(\gamma_{e0})S^{+}_{\gamma_{e}}(\omega_{0}).
\end{equation}
\end{widetext}

\section{Discussion}
\label{sec:discussion}
In this section we discuss the results of the calculations of the parameters characterizing qubit decoherence and we analyze a practical circuit which can be used for single-shot readout of a flux qubit.

We start with a discussion on the emission and absorption rates, along the lines of similar analysis done for charge~\cite{schoelkopf_2002_1} and charge-phase~\cite{martinis_2003_1} qubits. Both $\Gamma_{\downarrow}$ and $\Gamma_{\uparrow}$ are proportional to $\sin^{2}(\theta)$, due to the fact that the operator $\hat{\tau}_{x}$ (see \ref{eq:interaction_Hamiltonian_transformed}) causes transitions between the energy eigenstates.  The difference between the two rates given by \ref{eq:emission_rate} and \ref{eq:absorption_rate} is due to the last term in the integrand in expression \ref{eq:perturbation_expression}, connected with the fact that a commutator is non-vanishing, so it can be attributed to quantum noise. Let $p_{g}$ and $p_{e}$ be the probabilities for the qubit to be respectively in the ground and in the excited state. The time evolution of these probabilities is determined by the rate equations: 
\begin{eqnarray}\label{eq:evolution_probabilities}
\frac{dp_{g}}{dt}&=&-\frac{\Gamma_{\uparrow}}{2}p_{g}+\frac{\Gamma_{\downarrow}}{2}p_{e}\\\nonumber
\frac{dp_{e}}{dt}&=&\frac{\Gamma_{\downarrow}}{2}p_{g}-\frac{\Gamma_{\uparrow}}{2}p_{e}
\end{eqnarray}
and the normalization condition $p_{g}+p_{e}=1$. Since $\Gamma_{\downarrow}$ and $\Gamma_{\uparrow}$ in \ref{eq:emission_rate} and \ref{eq:absorption_rate} describe the decay of $\langle \hat{\tau}_{z}\rangle$, they appear divided by 2 in \ref{eq:evolution_probabilities}. The polarization $P(t)=p_{g}(t)-p_{e}(t)$ tends to the equilibrium value 
\begin{equation}\label{eq:stationary_polarization}
P_{s}=\frac{\Gamma_{\downarrow}-\Gamma_{\uparrow}}{\Gamma_{\downarrow}+\Gamma_{\uparrow}}
\end{equation}
with a relaxation rate
\begin{equation}\label{eq:relaxation_rate}
\Gamma_{r}=\frac{\Gamma_{\downarrow}+\Gamma_{\uparrow}}{2}.
\end{equation}

The spectral densities of the symmetrized and anti-symmetrized correlation functions \ref{eq:sym_corr_function} and \ref{eq:asym_corr_function} depend on the impedance at the port $P_{m}$ (see Section \ref{sec:response}) as:
\begin{equation}\label{eq:Splus}
S_{\gamma_{e}}^{+}(\omega)=\frac{8\pi}{\omega}\coth\left(\frac{\hbar\omega}{2k_{B}T}\right)\frac{Re(Z_{mm}(\omega))}{R_{K}}
\end{equation}
and
\begin{equation}\label{eq:Sminus}
-iS_{\gamma_{e}}^{-}(\omega)=\frac{8\pi}{\omega}\frac{Re(Z_{mm}(\omega))}{R_{K}},
\end{equation}
where $R_{K}=h/e^{2}$ (see Devoret~\cite{devoret_1995_1}).
Given the relations \ref{eq:relaxation_rate}, \ref{eq:emission_rate}, \ref{eq:absorption_rate} and \ref{eq:spectral_density} and the properties of $S_{\gamma_{e}}^{+}(\omega)$ and $S_{\gamma_{e}}^{-}(\omega)$ to be respectively even and odd functions, the relaxation rate can be written as 
\begin{equation}\label{eq:relaxation_rate_expression}
\Gamma_{r}=\frac{1}{2\hbar^{2}}\sin^{2}(\theta)k^{2}(\gamma_{e0})[S^{+}_{\gamma_{e}}(\omega_{01}+\omega_{0})+S^{+}_{\gamma_{e}}(\omega_{01}-\omega_{0})].
\end{equation}
For a flux qubit coupled to a DC-SQUID biased with a constant current van~der~Wal \emph{et al.}~\cite{vanderwal_2003_1} found that qubit relaxation is proportional to $S^{+}_{\gamma_{e}}(\omega_{01})$. Our results show that because of driving the SQUID with an AC bias current at frequency $\omega_{0}$ the qubit relaxation rate is proportional to the sum of $S^{+}_{\gamma_{e}}(\omega_{01}+\omega_{0})$ and $S^{+}_{\gamma_{e}}(\omega_{01}-\omega_{0})$, and multiplied by the coupling factor of Eq.~\ref{eq:coupling_factor}. In practice $\omega_{0}<<\omega_{01}$, which implies that for a spectral density of the noise which is reasonably flat at large frequencies we can take $S^{+}_{\gamma_{e}}(\omega_{01}+\omega_{0})\sim S^{+}_{\gamma_{e}}(\omega_{01}-\omega_{0})\sim S^{+}_{\gamma_{e}}(\omega_{01})$ and our results is not significantly different of the case of a DC current biased SQUID~\cite{vanderwal_2003_1}.

From \ref{eq:spectral_density}, \ref{eq:Splus} and \ref{eq:Sminus} it follows that
\begin{equation}\label{eq:spectral_density_ratios}
\frac{S_{\gamma_{e}}(-\omega)}{S_{\gamma_{e}}(\omega)} = e^{-\beta\hbar\omega},
\end{equation}
where $\beta=1/(k_{B}T)$. For a qubit coupled to a DC-SQUID biased with a constant current, which is the case analyzed by van~der~Wal \emph{et al.}~\cite{vanderwal_2003_1}, $\Gamma_{\downarrow}$ and $\Gamma_{\uparrow}$ are proportional to the spectral density of the noise at the frequencies $\omega_{01}$ and $-\omega_{01}$, respectively. For that case Eq. \ref{eq:spectral_density_ratios} implies that $\Gamma_{\uparrow}/\Gamma{\downarrow}= e^{-\beta\hbar\omega_{01}}$. This is the detailed balance condition and implies that in a stationary situation the qubit is in thermal equilibrium with the environment at temperature $T$ (see also the results of Schoelkopf \emph{et al.}~\cite{schoelkopf_2002_1}). In contrast, the relations \ref{eq:emission_rate} and \ref{eq:absorption_rate} show that for the case analyzed here the detailed balance condition is in general not satisfied.

The dephasing rate in \ref{eq:dephasing_rate} can be written, if \ref{eq:relaxation_rate} is used, as
\begin{equation}\label{eq:dephasing_rate_simple}
\Gamma_{\phi}=\frac{\Gamma_{\downarrow}+\Gamma_{\uparrow}}{4}+\Gamma_{\phi}^{*},
\end{equation}
where the \emph{pure dephasing rate} $\Gamma_{\phi}^{*}$ is given by
\begin{equation}\label{eq:pure_dephasing_rate}
\Gamma_{\phi}^{*}=\frac{1}{\hbar^{2}}\cos^{2}(\theta)k^{2}(\gamma_{e0})S^{+}_{\gamma_{e}}(\omega_{0}).
\end{equation}
The factor $\cos^{2}(\theta)$ is due to the coefficient of the operator $\hat{\tau}_{z}$ in \ref{eq:interaction_Hamiltonian_transformed}. Dephasing is a result of random modulation of energy level separation due to noise in the SQUID circulating current. The fact that the SQUID is driven with an AC current has the consequence that the pure dephasing rate depends on noise at $\omega_{0}$, which is qualitatively different of the result obtained by van~der~Wal \emph{et al.}~\cite{vanderwal_2003_1}. For the radio-frequency Bloch-transistor electrometer~\cite{zorin_2001_1} a similar contribution of the converted noise to back-action was found~\cite{zorin_2001_1,zorin_2002_1}. 

We compare the pure dephasing rate $\Gamma_{\phi}^{*}$ given by \ref{eq:pure_dephasing_rate} with a similar contribution due to the second term in Eq. \ref{eq:flux_backaction} as calculated by Shnirman~\emph{et al.}~\cite{shnirman_2002_1}, that we denote by $\widetilde{\Gamma}_{\phi}^{*}$. We consider the simple case where the DC-SQUID is shunted by a resistor $R_{sh}$, corresponding to ohmic dissipation, when the result of Shnirman~\emph{et al.} can be used. The following relation is valid for $\omega_{0}L_{J0}<<R_{sh}$:
\begin{equation}\label{eq:comparison_rates}
\frac{\Gamma_{\phi}^{*}}{\widetilde{\Gamma}_{\phi}^{*}}=\left(\frac{\hbar\omega_{0}}{k_{B}T}\right)^3\coth\left(\frac{\hbar\omega_{0}}{k_{B}T}\right)\frac{\gamma_{e0}^2}{2}\frac{R_{sh}R_{K}}{(\omega_{0}L_{J0})^2}.
\end{equation}
For the case $\omega_{0}L_{J0}<<R_{K}$ and $\hbar\omega_{0}\sim k_{B}T$, the dephasing rate $\widetilde{\Gamma}_{\phi}^{*}$ is dominant even at small SQUID driving amplitudes. 

The reliable measurement of the qubit state requires that the AC voltage at the output of the circuit is averaged for a long enough time, such that the noise due to the amplifier is less that the difference between the voltage values corresponding to the two qubit flux states. We define the discrimination time as the time necessary to have a measurement signal to noise ratio equal to 1. It is thus given by 
\begin{equation}\label{eq:discrimination_time}
T_{discr}=\frac{S_{V}(\omega_{0})}{(\Delta V_{qb})^2},
\end{equation}
where $S_{V}(\omega_{0})$ is the spectral density of the voltage noise and $\Delta V_{qb}$ is the difference in the output voltage values corresponding to the two qubit states. The value of $\Delta V_{qb}$ is proportional to $\gamma_{e0}$. The discrimination time $T_{discr}$, the relaxation time $T_{r}=1/\Gamma_{r}$ and the dephasing time $T_{\phi}=1/\Gamma_{\phi}$ are inversely proportional to $\gamma_{e0}^{2}$ (see \ref{eq:relaxation_rate} with \ref{eq:emission_rate} and \ref{eq:absorption_rate}, and \ref{eq:dephasing_rate}). Increasing the amplitude of the AC driving leads to a decrease in the discrimination time. However, this is accompanied by a proportional decrease of the qubit decoherence times $T_{r}$ and $T_{\phi}$. This illustrates the tradeoff between obtaining information about a quantum system and the state disturbance.
The measurement is efficient if the ratio $T_{r}/T_{discr}$ is large. The ratio $T_{r}/T_{discr}$ does not depend on the amplitude of the AC driving. However, a fast measurement is necessary if we take into account the fact that, besides the measurement back-action, there are also other sources of decoherence that will increase the total relaxation rate.

\begin{figure}[!]
\includegraphics[width=3.4in]{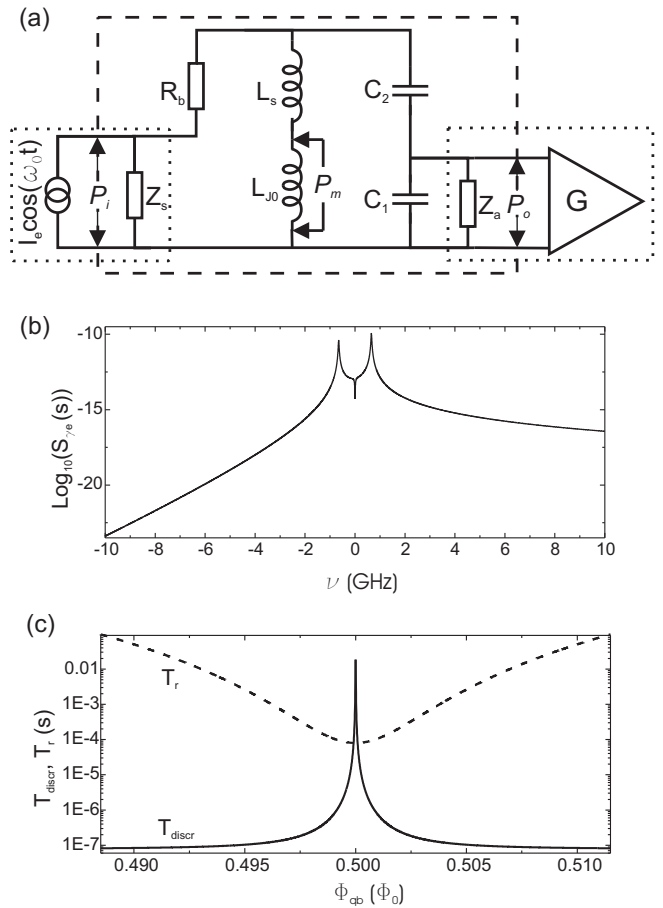}
\caption{\label{fig3} (a) Schematic representation of the measurement circuit, with notations according to Sec. \ref{sec:response}. The values of the circuit elements are $R_{b} = 4.7$ k$\Omega$, $Z_{s}=Z_{a}=50$ $\Omega$, $L_{J0}=1.76$ nH, $L_{s}=0.18$ nH, $C_{1}=60.7$ pF and $C_{2}=60.6$ pF. (b) Representation of the Fourier transform of the correlation function for the SQUID phase operator versus frequency. (c) Plot of the measurement discrimination time (continuous line) and qubit relaxation time $T_{r}$ (dashed line) versus qubit bias flux. The measurement is performed with an amplitude of the AC driving such that $\gamma_{e0}=0.5$ at a frequency $\nu_{0}=672$ MHz.} 
\end{figure}

We now analyze the measurement of a flux qubit using our particular SQUID embedding network presented in Fig.~\ref{fig3}a. The bias resistor $R_{b}$ has the purpose to increase the impedance of the current source. The inductor $L_{s}$ is a small stray contribution, unavoidable in the design of the circuit. The combination of the capacitors $C_{1}$ and $C_{2}$ is an impedance transformer that will increase the effective impedance of the amplifier input, at the cost of a division of the total voltage across the inductors; they also provide the capacitive part necessary to create a resonant circuit. The DC-SQUID has Josephson junctions with a critical current $I_{c0}=200$ nA. The external magnetic flux in the SQUID loop corresponds to $f_{x}=3.35$, resulting in a critical current $I_{c}=187$ nA. The measured persistent current qubit has $I_{p} = 300$ nA and $\Delta = 5.5$ GHz. Figure~\ref{fig1}a and b show plots of the energy eigenvalues and persistent current expectation value versus bias flux for these qubit parameters. If the mutual inductance between the qubit and the DC-SQUID is $M=40$ pH, the relative change in Josephson inductance is given by $\alpha=3.4\:\%$. A plot of the expression $S_{\gamma_{e}}(\omega)$ given by \ref{eq:spectral_density} with \ref{eq:Splus} and \ref{eq:Sminus} is shown in Fig.~\ref{fig3}b, assuming a temperature $T=30$ mK. 

To calculate the discrimination time given by Eq.~\ref{eq:discrimination_time}, we assume that the voltage noise is dominated by the voltage amplifier. We assume that a low noise cryogenic amplifier with a noise temperature of 4 K is used~\cite{bradley_1999_1}. Equations \ref{eq:voltage_up_converted} and \ref{eq:voltage_down_converted} allow the calculation of $\Delta V_{qb}$. We assume that $\omega \sim 0$, since qubit relaxation is slow compared to the detector bandwidth (which will be confirmed by our calculation of the relaxation time) and we choose the value $\omega_{0}=2\pi\times 672$ MHz that gives a maximum amplitude $\Delta V_{qb}=189$ nV for an AC driving current such that the amplitude of the SQUID phase oscillations is $\gamma_{e0}=0.5$. The discrimination time is plotted in Fig.~\ref{fig3}c. The discrimination time increases when the qubit bias flux $\Phi_{q}$ approaches $\Phi_{0}/2$, because the difference between the expectation values of the qubit current for the two energy eigenstates decreases (see Fig.~\ref{fig2}b). 

The relaxation time is calculated using the expression~\ref{eq:relaxation_rate_expression} and plotted in Fig.~\ref{fig3}b for the chosen operating frequency $\omega_{0}$. The relaxation time away from the symmetry point $\Phi_{q}=\Phi_{0}/2$ increases as a result of both the decrease in the transverse coupling term $\sin^{2}(\theta)$ and the decrease in the real part of the impedance $Z_{mm}$ away from the resonance peak (see Eq.~\ref{eq:relaxation_rate_expression} and \ref{eq:Splus}). Over a wide range of parameters the relaxation time is considerably higher than the discrimination time, which allows for very efficient readout of the qubit state. Using a SQUID amplifier~\cite{muck_2001_1} with a noise temperature less than $100$ mK would allow for the reducing the discrimination time by more than one order of magnitude.

The measurement of the qubit state can be performed by applying the AC current to the SQUID for a time $T_{m}$ and by measuring the average AC voltage during this time interval. Note that the readout does not have to be performed at the same qubit bias flux where qubit manipulation prior to measurement is performed. It is possible to perform operations on the qubit at $\Phi_{qb}=\Phi_{0}/2$, where the qubit is insensitive to magnetic flux fluctuations. Afterwards the flux in the qubit can be changed adiabatically to a value where the two energy eigenstates have sufficiently different values of the persistent current to allow discrimination by the measurement. If the measurement is performed at $\Phi_{qb}=0.497\:\Phi_{0}$ where $T_{discr}=150$ ns and $T_{r}=50\:\mu s$, a measurement time $T_{m}=500$ ns ensures a measurement fidelity larger than $80\:\%$ and the qubit relaxation is negligible during this time interval.

The dephasing time depends on the Fourier transform of the symmetrized correlation function at the frequency of the AC driving. It follows from ~\ref{eq:Splus} that $S_{\gamma_{e}}^{+}(\omega_{0})$ is large, because $\omega_{0}$ has to be close to the resonance frequency of the circuit for efficient state readout. Even a small amplitude of the AC signal can cause significant dephasing of the qubit. During qubit manipulation, when no measurement is performed, the SQUID AC driving current has to be suppressed very strongly. For operation at a qubit energy level splitting $\omega_{01}=2\Delta$ the decoherence time due to the SQUID is $10\:\mu s$ if the amplitude of the phase oscillations is $|\gamma_{e0}|=0.003$.  

The continuous nature of the flux detection makes this readout method suitable for fundamental studies of the dynamics of the measurement process. Further analysis will be necessary for understanding the dynamics of the coupled qubit-SQUID system and for an evaluation of possible direct observation of qubit coherent evolution, similar to the situation described by Korotkov and Averin~\cite{korotkov_2001_1}.

\section{Conclusions}
\label{sec:conclusion}
In this paper we analyzed the DC-SQUID in the inductive mode as a readout method for superconducting flux qubits. We characterized the response function of the DC-SQUID as a flux detector. We described the back-action of the measurement circuit on the qubit. The relaxation and dephasing rates are proportional to circuit noise at frequencies that are shifted by the SQUID AC driving frequency, which is a result qualitatively different of the case of a the measurement done with a switching DC-SQUID. For a realistic measurement circuit, we found that single shot measurement of a flux qubit is possible.

\begin{acknowledgments}
This work was supported by the Dutch Organization for Fundamental Research on Matter (FOM), the European Union SQUBIT-2 project, and the U.S. Army Research Office (grant DAAD 19-00-1-0548).
\end{acknowledgments}

\bibliographystyle{apsrev}

\end{document}